# Controllable perfect spatiotemporal optical vortices

Shuoshuo Zhang[1,2,†], Zhangyu Zhou[1,†], Zhongsheng Man[3,†], Jielei Ni[1], Changjun Min[1], Yuquan Zhang[1,*], and Xiaocong Yuan[1,4]

[1]Nanophotonics Research Center, Institute of Microscale Optoelectronics & State Key Laboratory of Radio Frequency Heterogeneous Integration, Shenzhen University, Shenzhen 518060, China.

[2]School of Physical Science and Technology, Northwestern Polytechnical University, Xi'an 710129, China.

[3]School of Physics and Optoelectronic Engineering, Shandong University of Technology, Zibo 255000, China.

[4]Research Centre for Frontier Fundamental Studies, Zhejiang Lab, Hangzhou 311100, China.

[†]These authors contributed equally to this work.

*Corresponding authors: yqzhang@szu.edu.cn

**Abstract:** Spatiotemporal optical vortices (STOVs), as a kind of structured light pulses carrying transverse orbital angular momentum (OAM), have recently attracted significant research interest due to their unique photonic properties. However, general STOV pulses typically exhibit an annular intensity profile in the spatiotemporal plane, with a radius that scales with the topological charge, limiting their potential in many applications. Here, to address this limitation, we introduce the concept of perfect spatiotemporal optical vortices (PSTOVs). Unlike STOV pulses, the intensity distribution of PSTOV wavepackets is nearly independent of the topological charge. We show that such wavepackets can be generated by applying the spatiotemporal Fourier transform to a Bessel-Gaussian mode in the spatiotemporal frequency domain. More importantly, the mode distribution of PSTOV wavepackets can be freely controlled by introducing azimuthal-dependent phase modulation, enabling conversion from a standard annular profile to arbitrary polygonal shapes. Finally, experimental results confirm the successful generation of these wavepackets. Our findings will expand the study of STOV pulses and explore their potential applications in optical communications, information processing, topological photonics, and ultrafast control of light−matter interactions.

## Introduction

Synthesizing ultrafast light pulses with tailored spatial and temporal properties has found wide applications in laser microfabrication (*1*), nonlinear optics (*2*), extreme physics (*3*), and molecular dynamics research (*4,5*). Traditionally, light pulses are treated as space-time separable solutions to Maxwell's equations, which allows us to describe and control their spatial and temporal behaviors separately. However, this separation is not always valid. In many cases, the spatial distributions of light pulses depend on the temporal coordinates, also known as spatiotemporal coupling (STC) (*6*). STC effects are abundant in ultrafast optics, such as pulse-front tilt and pulse-front curvature, typically arising from imperfect dispersion compensation and aberrations in optical systems, and are often considered as a nuisance that needs to be avoided (*7*). Although these effects may seem undesirable, they can nevertheless serve as a "hidden" degree of freedom in light structuring, providing more capabilities for generating exotic spatiotemporal light states (*8,9*).

Recently, fruitful endeavors have been devoted to developing novel technologies to synthesize light pulses with complex spatiotemporal correlations (*10–12*). In particular, spatiotemporal optical vortices (STOVs) have garnered considerable interest in some of recent studies (*13,14*). Unlike spatial optical vortices, STOV pulses are characterized by a helical phase structure and energy circulation in the spatiotemporal plane, thereby carrying an intrinsic transverse orbital angular momentum (OAM) (*15,16*). Due to their sophisticated field structures, earlier studies on STOV pulses were limited to just a few theoretical works (*17,18*). The existence of STOV structures was first observed during the self-focusing process of light pulses (*19*). However, the nonlinear generation of STOVs is uncontrollable. As an improvement, the controllable generation of STOVs in a linear manner using a 4f pulse shaper was reported (*20,21*). Since then, a series studies on STOV pulses have been conducted (*22–44*). The discovery of photonic transverse OAM could provide new dimensions for manipulating light and is expected to expand application scopes of OAM states. Despite these rapid advancements, there are still some limitations that need to be addressed from an application perspective. For instance, general STOV pulses exhibit a characteristic annular intensity profile in the spatiotemporal domain, with the radius strongly dependent on the topological charge. This property may pose challenges in applications that require the superposition of multiple vortices with different topological charges, such as optical communications (*45,46*) and information processing (*47,48*).

In this work, we address this limitation by introducing the concept of perfect spatiotemporal optical vortices (PSTOVs). In contrast to general STOV pulses, the intensity distribution of PSTOV wavepackets is almost independent of the topological charge. Theoretical analysis and experimental results reveal that these remarkable wavepackets can be generated by the spatiotemporal Fourier transform of a Bessel-Gaussian beam in the spatiotemporal frequency domain. Furthermore, we demonstrate the mode conversion of the PSTOV wavepacket, from a standard annular profile to arbitrary polygonal shapes, which opens up exciting new possibilities for manipulating and utilizing these novel light waveforms. Such controllable PSTOV wavepackets are promising for optical communications, high-dimensional information processing, space-time photonic topologies, as well as ultrafast light−matter interactions.

## Results
### Theoretical framework
The design of the considered PSTOV wavepacket is inspired by the concept of perfect optical vortices (POVs) introduced by Ostrovsky et al. (*49*). Akin to the formation of POV beams, a PSTOV wavepacket can be generated through the spatiotemporal Fourier transform of a Bessel-Gaussian (BG) beam in the spatiotemporal frequency domain (Fig. 1A). The complex field of the BG beam in the spatiotemporal frequency domain ($k_x$−$\Omega$ plane) can be expressed as:

$$\tilde{\psi}(k_x,\Omega) = \tilde{\psi}(r,\varphi) \propto J_\ell(\beta r)\exp\left(-\frac{r^2}{w_0^2}\right)\exp(i\ell\varphi), \tag{1}$$

where $r = \sqrt{(k_x/\Delta k_x)^2 + (\Omega/\Delta\omega)^2}$ and $\varphi = \tan^{-1}(\Delta\omega k_x/\Delta k_x \Omega)$ are the normalized polar coordinates in the $k_x$–$\Omega$ plane; $k_x$ is the spatial frequency in the $x$-direction, and $\Omega$ is the temporal frequency defined with respect to $\omega_0$; $J_\ell$ is the Bessel function of the first kind, where $\ell$ is the topological charge (or order); $w_0$ is the waist radius, and $\beta$ is a parameter that controls the radial distribution of the BG mode. The field in the $x$–$t$ plane can be calculated by a two-dimensional (2D) Fourier transform:

$$\psi(\rho,\phi) = \text{F}\cdot\text{T}\cdot\{\tilde{\psi}(r,\varphi)\} = \int_0^{2\pi}\int_0^\infty \tilde{\psi}(r,\varphi)\exp[i\rho r\cos(\varphi-\phi)]r\,dr\,d\varphi, \tag{2}$$

where $\rho = \sqrt{(x/\Delta x)^2 + (t/\Delta t)^2}$ and $\phi = \tan^{-1}(-\Delta t x/\Delta x t)$ are the normalized polar coordinates in the $x$–$t$ plane; $x$ is the spatial coordinate, and $t$ is the temporal coordinate; $\Delta x$ and $\Delta t$ are the characteristic widths in the $x$- and $t$-directions, which are inversely proportional to the spatial and temporal bandwidths $\Delta k_x$ and $\Delta\omega$, respectively. Notably, the coordinate normalization in Eqs. (1) and (2) is necessary because the spatial (spatial frequency) and temporal (temporal frequency) coordinates have different units.

The integration of Eq. (2) yields the following expression (see Supplementary Text S1 for detailed derivation)

$$\psi(\rho,\phi) \propto \exp\left(-\frac{\rho^2+\beta^2}{w_1^2}\right)I_\ell\left(\frac{2\rho\beta}{w_1^2}\right)\exp(i\ell\phi), \tag{3}$$

where $w_1 = 2/w_0$, and $I_\ell$ is the modified Bessel function of the first kind. The above equation is exactly the expression of a PSTOV wavepacket in the spatiotemporal domain, as it exhibits a mathematical form similar to that of POV beams in the spatial domain (*50, 51*).

The common strategy for synthesizing complex spatiotemporal light pulses, such as the PSTOV wavepacket in this work, relies on the use of a Fourier-transform pulse shaper (*8, 9*). A conceptual sketch of the system is illustrated in Fig. 1B. The pulse shaper typically consists of a pair of gratings, a pair of cylindrical lenses, and a spatial modulator (SLM), with all components separated by the focal length *f* of the cylindrical lenses, forming a 4*f* system. In this configuration, the location of the SLM can be regarded as the spatiotemporal spectrum plane, enabling the modulation of input pulses in the $k_x$–$\Omega$ domain (see Supplementary Text S2 for details). To generate the desired PSTOV wavepacket, the SLM is encoded with a phase hologram given by:

$$\varphi_{SLM}(r,\varphi) = \text{mod}\left\{2\pi\frac{r_0(\varphi)}{f_0}r + \ell\varphi,\ 2\pi\right\}, \tag{4}$$

where $(r, \varphi)$ are the previously defined normalized polar coordinates in the $k_x$–$\Omega$ plane. The above expression can be considered as the superposition of a digital axicon lens (first term) and a helical phase (second term). Akin to the generation of a BG beam in the spatial domain, this hologram can be utilized to produce a BG mode in the $k_x$–$\Omega$ plane. After a spatiotemporal Fourier transform, the PSTOV wavepacket in the $x$–$t$ plane will be generated in the far field of the pulse shaper. Furthermore, to make the generated wavepacket have a controllable mode distribution, a parameter $r_0(\varphi)$ is used to control the shape of the digital axicon lens, while the parameter $f_0$ is set as a constant. The freely shaped digital lens behaves as an "optical pen", which allows us to generate variously shaped PSTOV wavepackets.

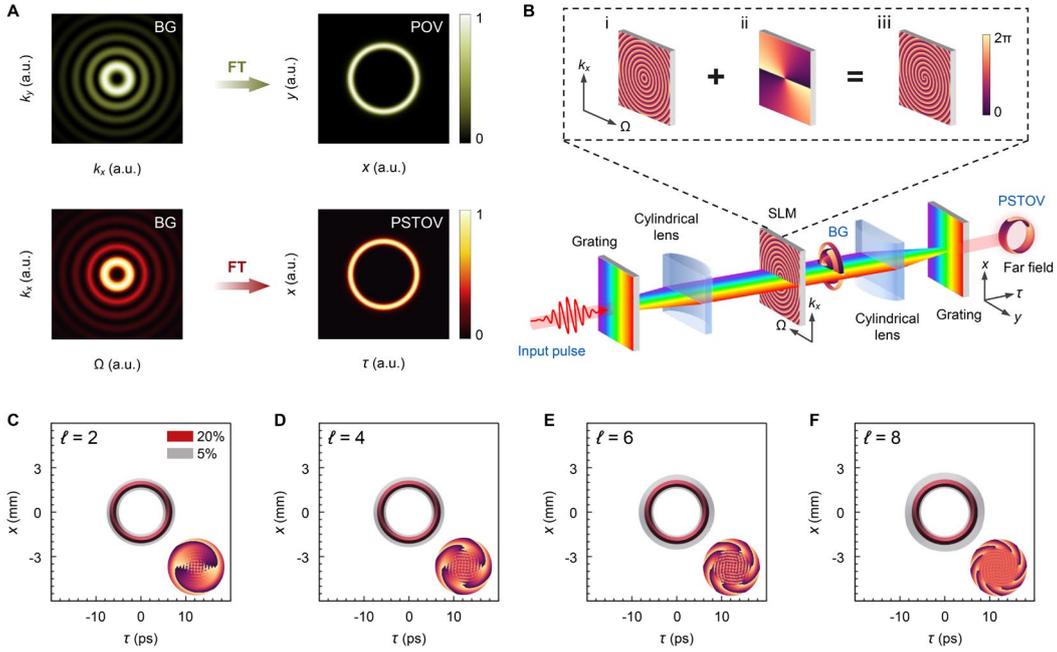

**Fig. 1. Principle of PSTOV wavepacket generation.** (**A**) Analogy between POV beams and PSTOV wavepackets. Akin to the formation of POV beams, a PSTOV wavepacket can be generated through the spatiotemporal Fourier transform (FT) of a Bessel-Gaussian (BG) beam in the spatiotemporal frequency domain ($k_x$–$\Omega$). (**B**) Concept of PSTOV wavepacket generation based on a 4f pulse shaper. A generic input pulse is first transformed into the $k_x$–$\Omega$ domain through a diffraction grating and a cylindrical lens, then modulated by a spatial light modulator (SLM) into a BG beam, and finally converted into a PSTOV wavepacket in the far field of the pulse shaper after a spatiotemporal Fourier transform. The inset shows the phase hologram encoded on the SLM, which can be derived by the superposition of (i) a digital axicon lens and (ii) a helical phase. (**C** to **F**) Simulated 3D iso-intensity profiles (20% and 5% of the maximum intensity) of the PSTOV wavepackets with topological charges $\ell$ = 2, 4, 6, and 8, respectively. Insets: spatiotemporal phase distributions in the $x$–$\tau$ plane. The numerical simulations are performed using the angular spectrum method (see Methods and Materials), with same parameters as those used in the experiment. The parameters are set as: $\lambda_0$ = 800 nm, $\Delta x$ = 0.15 mm, $\Delta t$ = 500 fs, $r_0$ = 0.5, $f_0$ = 0.125, $z$ = 0, and the local time coordinate $\tau = t - z / c$.

We first investigate the generation of annular PSTOV wavepackets, where $r_0(\varphi)$ is set as a constant.

The simulated 3D iso-intensity profiles the PSTOV wavepackets with topological charges $\ell$ = 2, 4, 6, and 8 are shown in Fig. 1 (C to F), respectively. The radius of the intensity distribution is nearly independent of the topological charge, similar to the POV beams in the spatial domain, demonstrating the feasibility of generating PSTOV wavepackets. The radius of the PSTOV wavepacket can be further controlled by the parameters $r_0$ and $f_0$. When a lager ratio of $r_0 / f_0$ is selected, a correspondingly larger radius is produced (see Supplementary Text S3 for details).

**Experimental results**

The experimental setup is illustrated in Fig. 2. In our experiment, the PSTOV wavepacket is generated using a 4$f$ pulse shaper and characterized by a Mach-Zehnder scanning interferometer. More experimental details are given in Materials and Methods. Since the refence pulse is considerably shorter than the generated PSTOV wavepacket, the spatiotemporal profile of the PSTOV wavepacket can be reconstructed from the delay-dependent interference patterns recorded by a charge-coupled-device (CCD) camera (see Supplementary Text S6 for details).

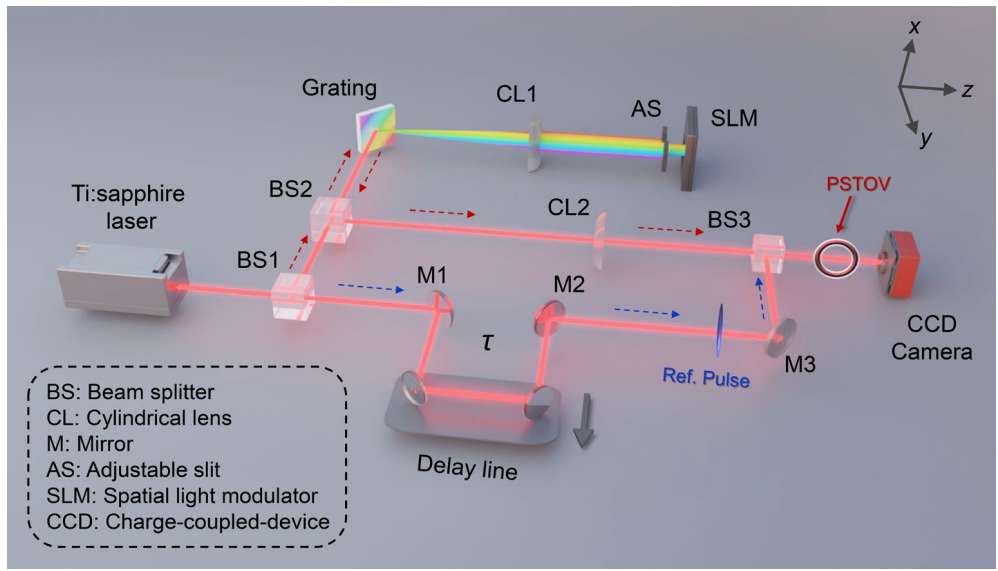

**Fig. 2. Experimental setup for generating and characterizing PSTOV wavepackets.** In the experiment, a nearly transform-limited input pulse from the laser source is split into two paths: in one path the PSTOV wavepacket is generated using a folded 4f pulse shaper consisting of a grating, a cylindrical lens (CL1), and a reflective spatial light modulator (SLM), while the other is the reference path. The generated PSTOV wavepacket is characterized using a Mach-Zehnder interferometer, where the reference path includes a motorized translation stage (Delay line). The interference patterns formed by the PSTOV wavepacket and the reference pulse are recorded by a charge-coupled-device (CCD) camera. An adjustable slit (AS) is placed before the SLM to tailor the temporal bandwidth (or pulse duration) of the input pulse.

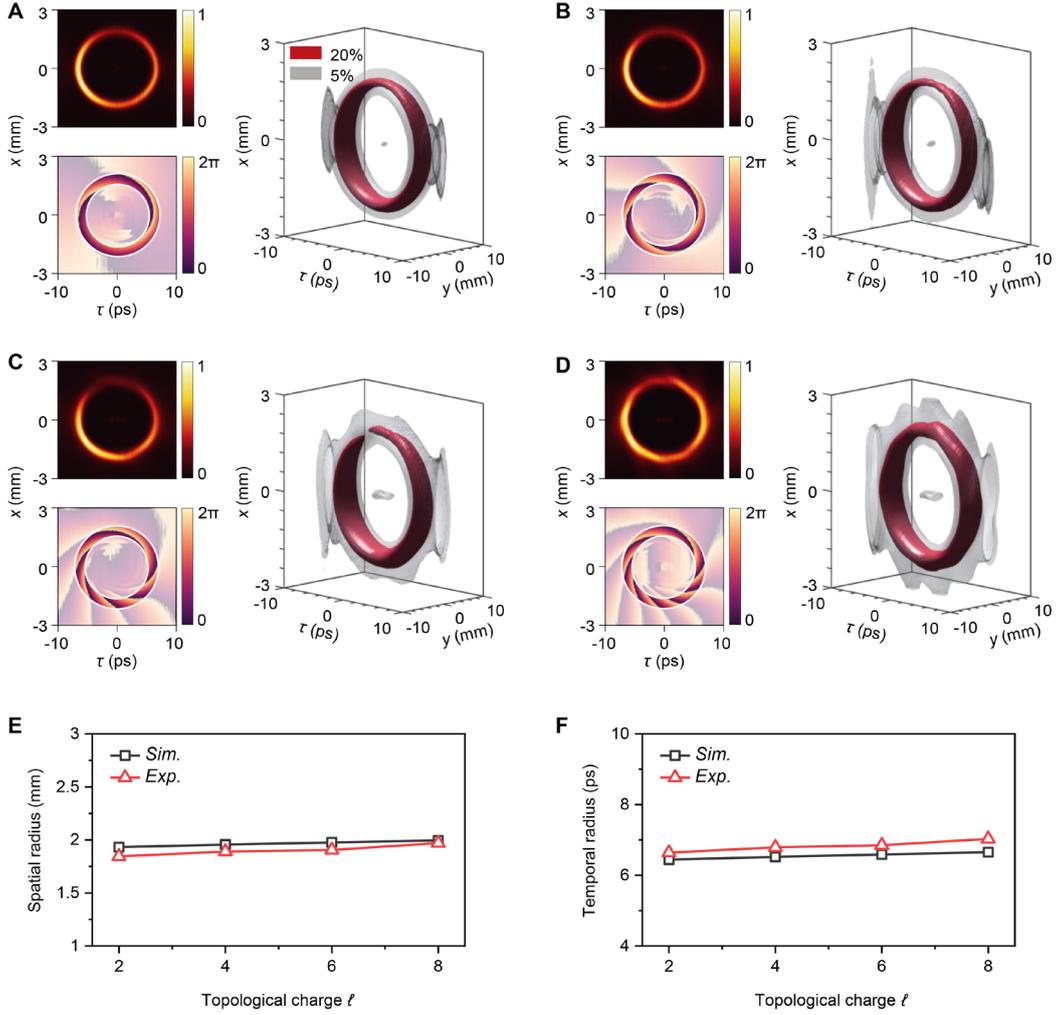

**Fig. 3. Experimentally generated PSTOV wavepackets.** (**A** to **D**) Measurements of the PSTOV wavepackets with topological charges $\ell$ = 2, 4, 6, and 8, respectively. Each panel shows the spatiotemporal intensity distribution in the $x-\tau$ plane (top left), the spatiotemporal phase distribution in the $x-\tau$ plane (bottom left), and the 3D iso-intensity profile (right) at 20% and 5% of the maximum intensity. The white circles in phase distributions highlight the helical phase structure of corresponding topological charge. (**E** and **F**) Dependence of spatial (E) and temporal (F) radii on the topological charge $\ell$. Sim.: simulated results; Exp.: experimental results.

Figure 3 (A to D) show the experimentally generated PSTOV wavepackets with topological charges $\ell$ = 2, 4, 6, and 8. The experimental results agree well with the simulated results depicted in Fig. 1 (C to F), except for the non-uniform intensity distribution and the slight distortion in the phase profiles. The non-uniform intensity distribution is primarily caused by the misalignment between the beam center and the phase hologram at the SLM plane. In addition, due to the limited diffraction efficiency of SLM, the unmodulated component of light forms the zero-order diffraction that is localized at the center of the generated PSTOV wavepackets. Compared to complex amplitude modulation, our pure-phase modulation approach offers high power usage. As a result, the intensity of the zero-order diffraction is much weaker than that of the PSTOV wavepackets. Furthermore, we analyzed the spatial and temporal

radii of these wavepackets. The simulated and experimental results as a function of topological charges are presented in Fig. 3 (E and F), indicating that their radii do not vary significantly with the topological charge.

**Propagation dynamics in free space**

Once PSTOV wavepackets are generated, it is important to understand how they propagate. To investigate the propagation dynamics, both theoretical and experimental analyses of two PSTOV wavepackets with opposite topological charges of $\ell = +4$ and $-4$ at different positions along the z-axis are conducted. The simulated intensities with $\ell = +4$ at $z = -200$ mm, 0, and 200 mm are shown in Fig. 4A. As illustrated, when the PSTOV wavepacket propagates forward in free space, its beam size along the x-direction gradually decrease, while the temporal scale remains unchanged, resulting in a variation in its eccentricity. Only at the Fourier plane ($z = 0$), a standard circularly symmetric wavepacket can be observed. This unique mode evolution can be attributed to the spatiotemporal astigmatism effect, which arises from the imbalance between spatial diffraction and material dispersion. Additionally, unlike general STOVs, which exhibit symmetric mode distributions about the Fourier plane, the PSTOV wavepackets show an asymmetric evolution. Notably, the results with $\ell = -4$ (Fig. 4D) are distinguishable as they resemble the mirror image of the $\ell = +4$ wavepacket. The experimental results shown in Fig. 4 (B and E) are in good agreement with simulated ones.

To further understand the physical mechanisms responsible for this unique mode evolution, we examine the energy density flux **J** of these two wavepackets (see Methods and Materials). As descried in Eq. (8), the energy density flux **J** scales with both the phase gradient and material dispersion. Due to the negligible dispersion in free space, there is no energy flow along the $\tau$-direction, as illustrate by the cyan and green arrows overlaid on the intensity distributions. Furthermore, the energy flow is tend to converge toward the center. This can qualitatively explain why the transverse beam size decreases with propagation. The OAM values at different propagating distances are calculated (see Methods and Materials). As shown in Fig. 4 (G and H), the OAM is quantized by $-\ell/2$ per photon and is conserved during propagation. The factor of 1/2 is a direct result of energy circulation restricted to $\pm x$, and the negative sign "$-$" arises from the reversal of topological charge caused by the spacetime coordinate transformation $\xi = -v_g\tau$.

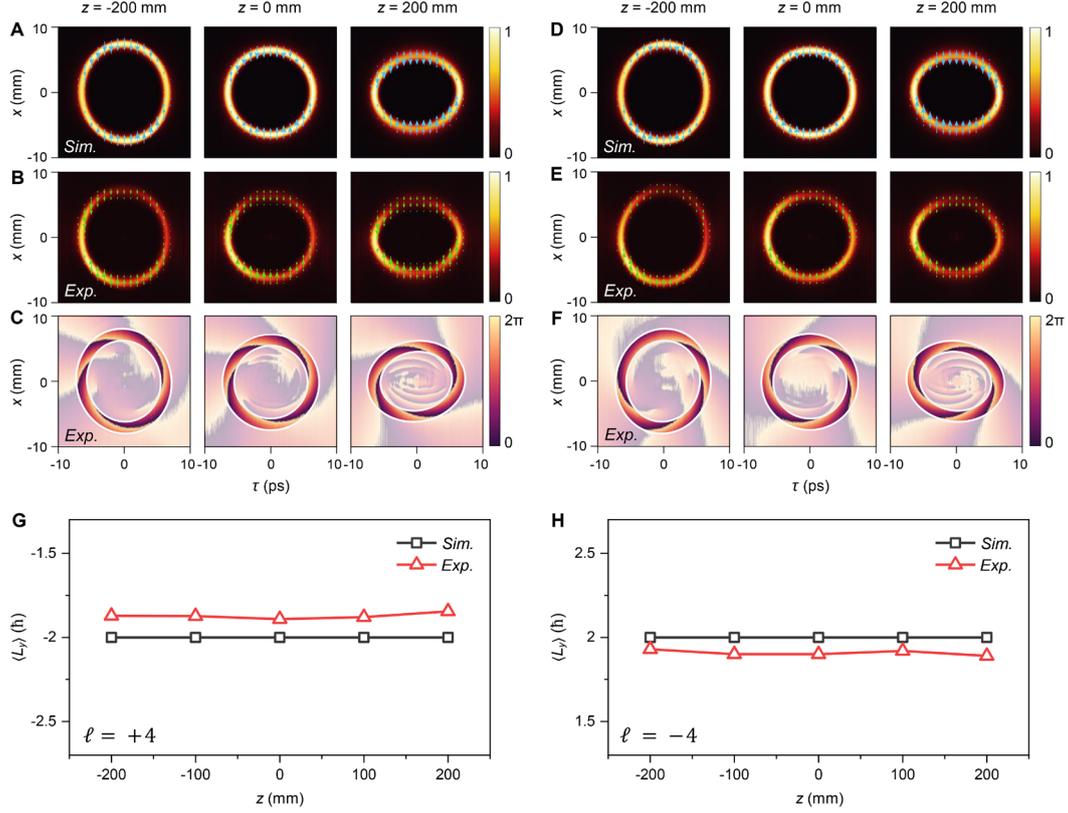

**Fig. 4. Propagation of PSTOV wavepackets.** (**A**) Simulated spatiotemporal intensity distributions of a PSTOV wavepacket with $\ell = +4$ at three different locations $z = -200$ mm, 0 mm, and 200 mm, showing the spatial evolution along longitudinal direction. (**B**) Experimental results for comparison with the simulations shown in (A). (**C**) Spatiotemporal phase distributions reconstructed from the experimental results shown in (B). (**D** to **F**) As in (A to C), but for a PSTOV wavepacket with $\ell = -4$. The cyan and green arrows in (A, B, D, and E) indicate the direction of the energy density flux **J**, computed using Eq. (8) (see Methods and Materials). (**G** and **H**) Propagation evolution of the OAM values for the two PSTOV wavepackets with $\ell = \pm 4$. Sim.: simulated results; Exp.: experimental results. The Rayleigh distance $z_R = k_0 \Delta x^2 / 2 \approx 88.4$ mm.

## Controllability of mode distribution

In the previous discussion, we focus on the generation of annular PSTOV wavepackets. In this section, let us demonstrate the ability to generate PSTOV wavepackets with more complex shapes. For differentiation purposes, we will refer to these as generalized PSTOV wavepackets. For example, a generalized PSTOV wavepacket with controllable polygonal shapes can be obtained by setting the parameter $r_0(\varphi)$ in Eq. (4) as follows:

$$r_0(\varphi) = 0.5 - \sin(q\varphi)/p, \tag{5}$$

here, $p$ controls the smoothness, while $q$ governs the shape of the generalized PSTOV wavepackets (see Supplementary Text S8 for details). By selecting appropriate parameters $(p, q)$ according to Eq. (5), the generalized PSTOV wavepackets with triangular ($p = 30$, $q = 3$), square ($p = 40$, $q = 4$), and pentagonal

($p$ = 50, $q$ = 5) shapes are generated experimentally, as shown in Fig. 5. Slight variation for $\ell$ = 2 and 6 indicates that the generalized PSTOV wavepackets share similar characteristics with annular PSTOV wavepackets. Finally, the 1D intensity curves along the $x$- and $\tau$-directions for the two squared PSTOV wavepackets with $\ell$ = 2 and $\ell$ = 6 are plotted in Fig. 5 (E and F). The highly overlapped curves confirm a weak correlation between the intensity distribution and the topological charge.

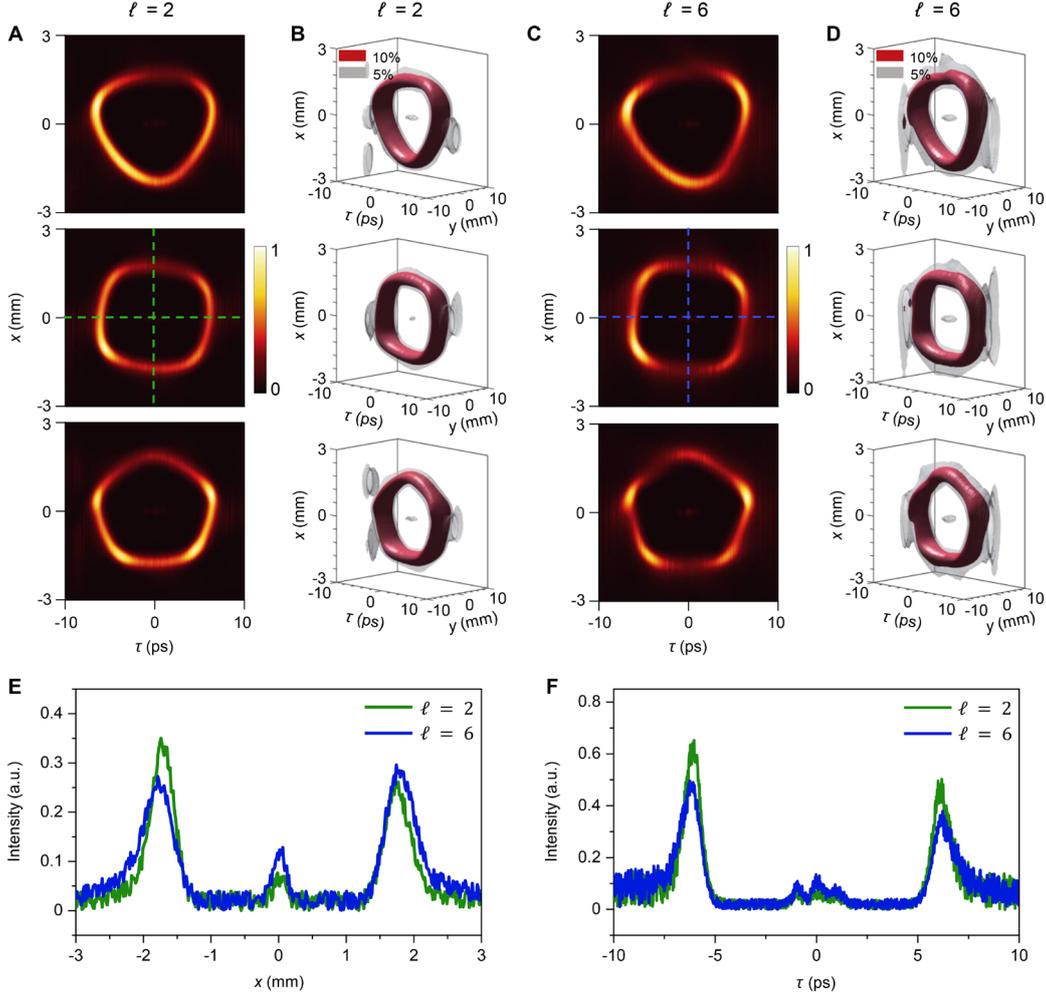

**Fig. 5. Generalized PSTOV wavepackets with controllable distributions.** (**A** and **B**) Measured spatiotemporal intensity distributions (A) and 3D iso-intensity profiles (B) of the generalized PSTOV wavepackets with triangular (top), square (medium), and pentagonal (bottom) mode distributions. Each with a topological charge $\ell$ = 2. (**C** and **D**) As in (A and B), but with a topological charge $\ell$ = 6. (**E** and **F**) Comparison of the 1D cross-sections of the intensity along the dotted lines in (A) and (C), showing the "perfectness" of generalized PSTOV wavepackets.

**Conclusion**

In conclusion, we have demonstrated the experimental generation of PSTOV wavepackets. Unlike previously reported STOV pulses, the PSTOV wavepackets exhibit unique features, including topological-charge-insensitive intensity distribution, asymmetric propagation evolution in free space, and controllable mode distribution. Theoretically, the PSTOV wavepackets can be generated by applying the spatiotemporal Fourier transform to a BG mode in the spatiotemporal frequency domain. The synthesis of these wavepackets relies on a Fourier-transform pulse shaper. Due to the geometric and physical differences from general STOV pulses, PSTOV wavepackets may find important applications in optical communications, information encryption, topological photonics, and ultrafast control of light−matter interactions

**Materials and Methods**

**Experimental setup**

The experimental configuration (Fig. 2) begins with a Ti:sapphire amplifier (Legend Elite HE, Coherent), which generates nearly transform-limited femtosecond pulses with a central wavelength of ~800 nm and a pulse duration of ~120 fs (see Supplementary Text S5 for details). The input pulses are then split into two paths by a beam splitter. One path serves as the reference arm, and the other is utilized to generate the PSTOV wavepacket via a folded 4f pulse shaper. The pulse shaper consists of a grating (1200 lines/mm), a cylindrical lens (f = 300 mm on the $y$-axis), and a reflective spatial light modulator (SLM; X15213-02, Hamamatsu). The SLM is placed at the focal plane of the cylindrical lens, allowing us to modulate the input pulses in the spatiotemporal frequency domain. After modulation, spatially dispersed frequencies will be reflected back and recombined at the grating. This process can be understood as a temporal Fourier transform ($\Omega \to t$). To generate the PSTOV wavepacket, an additional spatial Fourier transform ($k_x \to x$) is required. To achieve this, the modulated pulses are passed through a second cylindrical lens (f = 850 mm on the $x$-axis). The target PSTOV wavepacket is generated in the focal plane of the second lens (denoted as $z = 0$). Note that a standard PSTOV wavepacket should have a circularly symmetric intensity profile when projected onto the $x-z$ plane (i.e., $c\Delta t / \Delta x = 1$), where $\Delta x$ and $\Delta t$ represent the characteristic widths of the input pulses in the $x$- and $t$-directions, respectively. To match the spatial width $\Delta x$ (which is measured to be ~0.15 mm at $z = 0$), the temporal width $\Delta t$ is reshaped to ~500 fs by an adjustable slit before the SLM.

Finally, to characterize the generated fields, we use a Mach-Zehnder scanning interferometer to analyze the interference patterns formed by the PSTOV wavepacket and the reference pulse (see Supplementary Text S6 for details). The reference path contains a high-precision motorized translation stage (Delay line; V-408, PI). Therefore, the time delay $\tau$ between the two pulses can be adjusted. By scanning the time delay continuously, the full spatiotemporal profile of the generated PSTOV wavepacket can be reconstructed from the delay-dependent interference patterns recorded by a charge-coupled-device (CCD) camera. In our experiment, the scan step of the translation stage is set to 3 μm, corresponding to a time delay of ~20 fs.

**Numerical simulations**

The angular spectrum method is utilized to simulate the generation and propagation of PSTOV wavepackets in free space. Without loss of generality, we restrict our analysis to scalar pulses in the paraxial regime, and write the electric field $E(x, z; t) = \psi(x, z; t)\exp(ik_0 z - i\omega_0 t)$ as the product of a slowly-varying envelope $\psi(x, z; t)$ and a carrier wave of frequency $\omega_0$. The envelope function can be expressed as (8):

$$\psi(x, z; t) = \int_{-\infty}^{\infty}\int_{-\infty}^{\infty} \tilde{\psi}(k_x, \Omega) \exp\{i[k_x x + (k_z - k_0)z - \Omega t]\} dk_x d\Omega, \qquad (6)$$

where $\tilde{\psi}(k_x, \Omega)$ denotes the 2D Fourier transform of $\psi(x, 0; t)$, $k_x$ and $k_z$ are the spatial frequencies in the *x*- and *z*-directions, respectively; $\Omega = \omega - \omega_0$ is the temporal frequency defined relative to $\omega_0$, and $k_0 = \omega_0 / c$. For simplicity, the electric field is assumed uniform along the *y*-direction (i.e., $k_y = 0$), and therefore $k_z = \sqrt{(\omega/c)^2 - k_x^2}$.

To generate the PSTOV wavepacket, the input pulse is phase modulated in the spatiotemporal frequency domain. Accordingly, the modulated field can be expressed as:

$$\tilde{\psi}(k_x, \Omega) = \tilde{\psi}_G(k_x, \Omega) \exp[i\phi_{SLM}(k_x, \Omega)], \qquad (7)$$

where $\tilde{\psi}_G(k_x, \Omega) = \exp[-(k_x/\Delta k_x)^2 - (\Omega/\Delta\omega)^2]$ represents the spectrum of the input pulse, and $\phi_{SLM}(k_x, \Omega)$ is the phase hologram encoded on the SLM, as described by Eq. (4). For a transform-limited input pulse, its spatial and temporal bandwidths can be determined by $\Delta k_x = 2/\Delta x$ and $\Delta\omega = 2/\Delta t$, where $\Delta x$ and $\Delta t$ denote the pulse widths in the *x*- and *t*-directions, respectively. To reveal the propagation dynamics of the PSTOV wavepacket, we solved Eq. (6) with same parameters as those used in the experiment. Furthermore, since the PSTOV wavepacket propagates at a group velocity $v_g = c$ in free space, the wavepacket is characterized in a moving reference frame by introducing the local time coordinate $\tau = t - z/c$.

**Energy density flux and OAM calculation**

The energy density flux related to the complex envelope $\psi = |\psi|\exp(i\phi)$ can provide an intuitive picture of the dynamics of PSTOV wavepackets, especially when viewed together with propagation simulations. In the moving frame of the wavepacket, it can be derived as (52):

$$\mathbf{J} = k_0^{-1} |\psi|^2 \left( \nabla_\perp \phi - k_0 k_0'' \frac{\partial \phi}{\partial \tau} \hat{\tau} \right), \qquad (8)$$

where $\nabla_\perp$ denotes the transverse gradient operator, and $k_0'' = \partial^2 k/\partial\omega^2 (\omega = \omega_0)$ is the group velocity dispersion (GVD) parameter. It is clear from Eq. (8) that the transverse (*x* and *y*) components are proportional to the intensity and the phase gradient along the corresponding direction, while the longitudinal ($\tau$) also depends on the GVD parameter of the propagation medium.

The helical phase structure and associated energy flux in the $x-\tau$ plane give rise to a transverse OAM

along the *y*-direction. To evaluate the transverse OAM carried by the PSTOV wavepacket, it is crucial to determine the transverse OAM operator $\hat{L}_y$. As derived in Ref. (*15*), the operator $\hat{L}_y = -i(\xi\partial/\partial x + x\beta_2\partial/\partial \xi)$, where $\xi = -v_g\tau$ is a time-like longitudinal space coordinate, and $\beta_2 = v_g^2 k_0 k_0''$ is a dimensionless GVD parameter. Given the negligible dispersion in free space, we take $\beta_2 = 0$ and $v_g = c$ for all analyses. The average transverse OAM per photon, in units of $\hbar$, can be further calculated as:

$$\langle L_y \rangle = \frac{\langle \psi | \hat{L}_y | \psi \rangle}{\langle \psi | \psi \rangle} = \frac{\iiint \psi^* \hat{L}_y \psi \, \mathrm{d}x\mathrm{d}y\mathrm{d}\xi}{\iiint \psi^* \psi \, \mathrm{d}x\mathrm{d}y\mathrm{d}\xi}, \tag{9}$$

The origin of the spacetime coordinates (*x*, *y*, *ξ*) is chosen at the center of energy, so that the OAM is purely intrinsic.

**Funding:** This work was supported by the Guangdong Major Project of Basic and Applied Basic Research (2020B0301030009), National Natural Science Foundation of China (62375177, 62175157, and 12474294), Shenzhen Science and Technology Program (JCYJ20210324120403011 and RCJC20210609103232046), and Research Team Cultivation Program of Shenzhen University (2023QNT014).

**Author contributions:** S.Z. proposed the original idea. S.Z. and Z.Z. conceived and conducted the experiment. S.Z. and Z.M. performed the numerical simulations. S.Z. and Z.M. prepared the manuscript. J.N., C.M., Y.Z., and X.Y. revised the manuscript. Y.Z. guided and supervised the overall project. All authors analyzed the data and discussed the results.

**Competing interests:** The authors declare that they have no competing interests.

**Data and materials availability:** All data needed to evaluate the conclusions in the paper are present in the paper and/or the Supplementary Materials.